\documentclass[sigconf]{acmart}

\AtBeginDocument{%
  \providecommand\BibTeX{{%
    \normalfont B\kern-0.5em{\scshape i\kern-0.25em b}\kern-0.8em\TeX}}}

\copyrightyear{2024}
\acmYear{2024}
\setcopyright{rightsretained}
\acmConference[WWW '24] {Companion Proceedings of the ACM Web Conference 2024}{May 13--17, 2024}{Singapore, Singapore.}
\acmBooktitle{Companion Proceedings of the ACM Web Conference 2024 (WWW '24), May 13--17, 2024, Singapore}
\begin{document}

\title{Contextualizing Internet Memes Across Social Media Platforms}

\author{Saurav Joshi}
\affiliation{%
  \institution{University of Southern California} 
  \department{Information Sciences Institute}
  \city{Marina del Rey}
  \state{CA} 
  \country{USA}}
\email{syjoshi@isi.edu}

\author{Filip Ilievski}
\affiliation{%
  \institution{Vrije Universiteit Amsterdam}
  \country{The Netherlands}}
\email{f.ilievski@vu.nl}

\author{Luca Luceri}
\affiliation{%
  \institution{University of Southern California} 
  \department{Information Sciences Institute}
  \city{Marina del Rey}
  \state{CA} 
  \country{USA}}
\email{lluceri@isi.edu}

\renewcommand{\shortauthors}{Joshi, Ilievski, and Luceri}

\begin{abstract}
Internet memes have emerged as a novel format for communication and expressing ideas on the web. Their fluidity and creative nature are reflected in their widespread use, often across platforms and occasionally for unethical or harmful purposes. While computational work has already analyzed their high-level virality over time and developed specialized classifiers for hate speech detection, there have been no efforts to date that aim to holistically track, identify, and map internet memes posted on social media.
To bridge this gap, we investigate whether internet memes across social media platforms can be contextualized by using a semantic repository of knowledge, namely, a knowledge graph. We collect thousands of potential internet meme posts from two social media platforms, namely Reddit and Discord, and develop an extract-transform-load procedure to create a data lake with candidate meme posts. By using vision transformer-based similarity, we match these candidates against the memes cataloged in IMKG --- a recently released knowledge graph of internet memes. 
We leverage this grounding to highlight the potential of our proposed framework to study the prevalence of memes on different platforms, map them to IMKG, and
provide context about memes on social media.
\end{abstract}






\maketitle

\section{Introduction}
An \textit{internet meme (IM)} is ``a piece of culture, typically a joke, which gains influence through online transmission''~\cite{davison2012language}.
IMs have emerged in the digital age as a novel format for expressing ideas on the web, evolving from a simple form of entertainment to a sophisticated medium of communication that reflects societal sentiments and trends~\cite{bauckhage2013mathematical,naaman2011hip,bennett2003internet}. As they permeate almost every corner of our online world, there is a need to understand the depth and breadth of their influence~\cite{luceri2018social,thakur2023nesy}. The fluidity and relatable nature of IMs are reflected in their widespread use across both mainstream and fringe platforms~\cite{zannettou2018origins}, while their succinctness and multimodality make memes a challenging puzzle for state-of-the-art computational systems~\cite{imkg}. The need to automatically understand IMs is emphasized by their non-negligible use for unethical or harmful purposes, causing moderation concerns and negatively impacting the democracy on the Web~\cite{thakur2023nesy,luceri2021social}.

Existing computational work on IMs on social media tracked their virality by analyzing their spread over time~\cite{marino2015semiotics,taecharungroj2014effect,ling2021dissecting}, without considering the semantics of the memes. Other work focuses on hate speech and misogyny identification~\cite{kiela2021hateful,fersini2022semeval} aiming at classifying the intent of a meme into harmful or not, yet they do so while isolating each meme from its original creation and posting context. Recent work~\cite{10179315} bridges these two directions by combining offensive properties and adaptations over time, but it does not include background knowledge about the meme origin, entities, and related properties.
As a result, the generalizability of such approaches has been proven to be problematic~\cite{kirk2021memes}. A complementary thrust focuses on creating resources of internet memes in the form of web encyclopedias or knowledge graphs (KGs)~\cite{imkg}. However, these resources have not been used to study the natural evolution of memes or their harmfulness. In summary, there is a lack of a holistic approach to tracking, mapping, and characterizing internet memes that can be applied to memes on social media platforms.


To bridge the gap between these parallel streams of research on memes, we investigate \textit{whether internet memes across social media platforms can be contextualized by using a semantic repository of knowledge, i.e., a knowledge graph}. We collect thousands of potential internet meme posts from two popular social media platforms, Reddit and Discord, and perform an extract-transform-load procedure to create a data lake with candidate memes. We encode these potential memes with a Vision Transformer (ViT) ~\cite{dosovitskiy2021image}, and match them based on visual similarity against the meme repository in the Internet Meme Knowledge Graph (IMKG)~\cite{imkg}. We formulate the following four research questions to delve into the intricacies of meme culture, focusing on the effectiveness of the IMKG in categorizing and understanding memes within social media platforms:

\begin{itemize}
    \item[\textbf{RQ1:}] \textit{Can we reliably map social media memes to IMKG?} 
    While memes are multimodal media in a visual format, not every visual item posted on social media is a meme. Distinguishing memes from non-memes is often a subtle goal, especially when the image contains a textual caption~\cite{sherratt2022towards}. 
    We expect that by leveraging IMKG, we can detect, identify, and understand memes with high precision and relatively high recall. 

    \item[\textbf{RQ2:}] \textit{How prevalent are IMKG memes on Reddit and Discord, and which communities are most populated with memes?} 
    With the plethora of IMs distributed across these platforms, it is crucial to identify subreddits and channels that are richest in terms of memes. Discerning such nuances in the IM landscape is valuable for various stakeholders, including social science researchers studying digital culture, content curators, moderators tracking content, and marketers aiming to leverage the potential of IMs. 

    \item[\textbf{RQ3:}] \textit{Which memes are most popular on Reddit and Discord?} 
    Popular memes can shed light on topics and themes that resonate within online user communities, enabling us to decode their shared interests and sentiments. Addressing RQ3, we offer a snapshot of our framework's potential to analyze evolving trends and related community engagement. 
    

    \item[\textbf{RQ4:}] \textit{How does IMKG aid in a more refined comprehension of meme culture?} 
    IKMG's connections between meme templates, original media frames, and IM instances created by social media users, as well as its coverage of meme metadata and extracted entities, provide rich information that can possibly be exploited to contextualize social media memes, regardless of their origin. To gain insight into RQ4, we perform a qualitative analysis of the value brought by IMKG in understanding and characterizing social media memes.
\end{itemize}
We provide evidence that memes published online can be identified by mapping them to IMKG. We leverage this grounding to study the prevalence of memes on Reddit and Discord, identify popular memes, and find common meme channels and subreddits. Then, we showcase how the grounding enriches the meme context thanks to their link to the KG. In summary, our contributions are: \textit{(i)} To the best of our knowledge, we are the first to investigate the possibility of identifying 
memes on social media platforms by mapping them to a knowledge graph. To do so, we develop a comprehensive methodology that consists of meme source identification, meme candidate extraction, and grounding-driven meme identification; \textit{(ii)} We realize this methodological framework by collecting a dataset from two platforms, identifying memes, and obtaining further information about them from the IMKG; \textit{(iii)} We formulate four well-motivated research questions and leverage our methodological framework to answer them. Our experiments highlight the value of grounding memes ``in the wild'' to a central knowledge repository. To enable related research directions, we release the entire code and data from our analysis.\footnote{\url{https://github.com/usc-isi-i2/social-media-meme-identification}}



\section{Related Work}
\label{sec:relatedwork}

Prior work on analyzing internet memes has focused either on their propagation and virality, or on their semantics. We review both streams of work in this section.

\textbf{Propagation and virality.} Previous research on IMs within the realm of AI predominantly concentrates on understanding their propagation and virality on social media platforms over time. Social media platforms foster social interactions centered around user-generated content among individuals, organizations, and groups~\cite{mills2012virality,luceri2020measurement}. The shift from broadcasting (one-to-many) communication to social dialogues (many-to-many) is evident within these networks~\cite{kilian2012millennials}, transforming passive audiences into active authors~\cite{harlow2013facebook}, including malicious actors spreading harmful content \cite{ezzeddine2023exposing,luceri2021down}. Notably, a subset of these studies specifically delves into the communication dynamics of humorous memes on digital platforms~\cite{taecharungroj2015humour,vasquez2019language}. These works suggest that humor may significantly influence meme spread, fostering user interaction and fortifying a sense of community among users. Several studies have accurately forecasted the virality of IMs using social network indicators~\cite{weng2014predicting}. Furthermore, researchers have developed mathematical models that mirror the actual dissemination patterns of memes across the Internet~\cite{weng2012competition,wang2011epidemiological}. Beyond social network indicators, meme content and its presentation can also impact its popularity. Thus,~\cite{tsur2015don} examined the characteristics that contribute to the popularity of memes on Twitter, particularly focusing on their hashtags. Moreover, they highlighted the role of specific readability features, such as capitalization, in influencing a meme's popularity. This notion becomes especially relevant when considering the challenges in detecting hateful memes, which often adapt and evolve by intertwining harmful meanings with other cultural symbols or ideas. In~\cite{barnes2021dank}, the authors employed machine learning models to forecast the popularity of memes on Reddit and elucidate the underpinnings of meme virality. Complementing this, \cite{deza2015understanding} conducted experiments with Reddit data, introducing five cardinal visual attributes linked with the virality of IMs, namely: Animal, Synthetically Generated, (Not) Beautiful, Explicit, and Sexual. They subsequently developed a machine learning model that accurately predicts an image's viral potential. Complementary to these works that primarily focused on the virality and evolution of IMs on social media, our research provides a framework that uses a KG as a semantic source to ground IMs found on platforms like Reddit and Discord. We leverage this framework to explore the main sources, dissemination, and prevalence of memes on these platforms, while simultaneously providing background information about their origin and depicted entities.



\textbf{Meme semantics.} Understanding the underlying semantics of IMs is crucial not only to comprehend their cultural resonance but also to discern their potential misuse for propagating hateful, racist, misogynistic, and false content. \cite{park2020understanding} underscores the multifaceted contexts wherein users leverage memes, portraying them as potent instruments for emotional expression, particularly within intimate social media circles. In a broad study spanning 13 months across major web communities, it has been observed that memes, due to their vast propagation and evolution, play a pivotal role in influencing public opinion, and their proliferation is particularly pronounced in fringe communities where racist themes are prevalent \cite{zannettou2018origins}. The findings of \cite{das2020detecting} underscore that memes also possess the potential for misuse in spreading hate speech targeting specific societal groups based on factors like race, religion, and other identifiers. As pinpointed by \cite{kiela2020hateful}, the challenge lies in effectively detecting such harmful content amidst the vast collection of benign memes. Given the IM scale, manual monitoring on platforms like Reddit, Facebook, or Discord is nearly untenable. The complexities and open research issues in employing both textual and visual information for meme analysis are reviewed in \cite{afridi2021multimodal}. Hence, the need for AI-driven, automated solutions becomes increasingly important, a theme explored further in \cite{alayrac2022flamingo,singh2022flava,10179315}. \cite{suryawanshi2020multimodal} introduced the MultiOFF dataset, specifically curated for offensive content detection in memes, and demonstrated that combining image and text modalities using an early fusion technique outperforms single-modality baselines in identifying offensive content, addressing the implicit and multimodal nature of humor and sarcasm in memes. Additionally, \cite{chen2022multimodal} introduced a vision-language model that improves the performance of deep learning-based detection of hateful memes by involving a more substantial alignment between the text caption and visual information. To enable users to interpret the model behavior, Thakur et al. \cite{thakur2023nesy} experiment with example- and prototype-based methods for interpretable meme classification. \cite{cao2023prompting} explores a prompting-based method for classifying hate speech in memes, demonstrating the potential of utilizing prompts to enhance the detection and classification of hateful content within memes. Moreover, MemeGraphs~\cite{kougia2023memegraphs} introduces a novel method for meme classification utilizing scene graphs and knowledge graphs to provide structured representations for processing multimodal documents.
Crucial to addressing these challenges is developing and using robust semantic resources that enable deeper meme understanding. Such resources exist on the Web: for instance, KnowYourMeme (KYM) is a catalog that documents emerging viral Web phenomena, which often revolve around IMs.\footnote{\url{https://knowyourmeme.com}} 
The Internet Meme Knowledge Graph (IMKG)~\cite{imkg} is a comprehensive database that combines meme encyclopedias, meme generators, and general-purpose KGs to capture the expansive realm of IMs.
IMKG's approach merges semantic layers with textual, visual, and social subtleties, offering an in-depth perspective of meme culture.
Yet, IM resources like IMKG have not been leveraged in prior work to identify and characterize memes posted on social media. Our present paper aims to bridge this gap and provides an opening for future work to leverage this grounding to build generalizable models for hate speech and misinformation.

\section{Methodology}
\label{sec:method}

\begin{figure*}[ht!]
    \centering
    \includegraphics[width=0.75\textwidth]{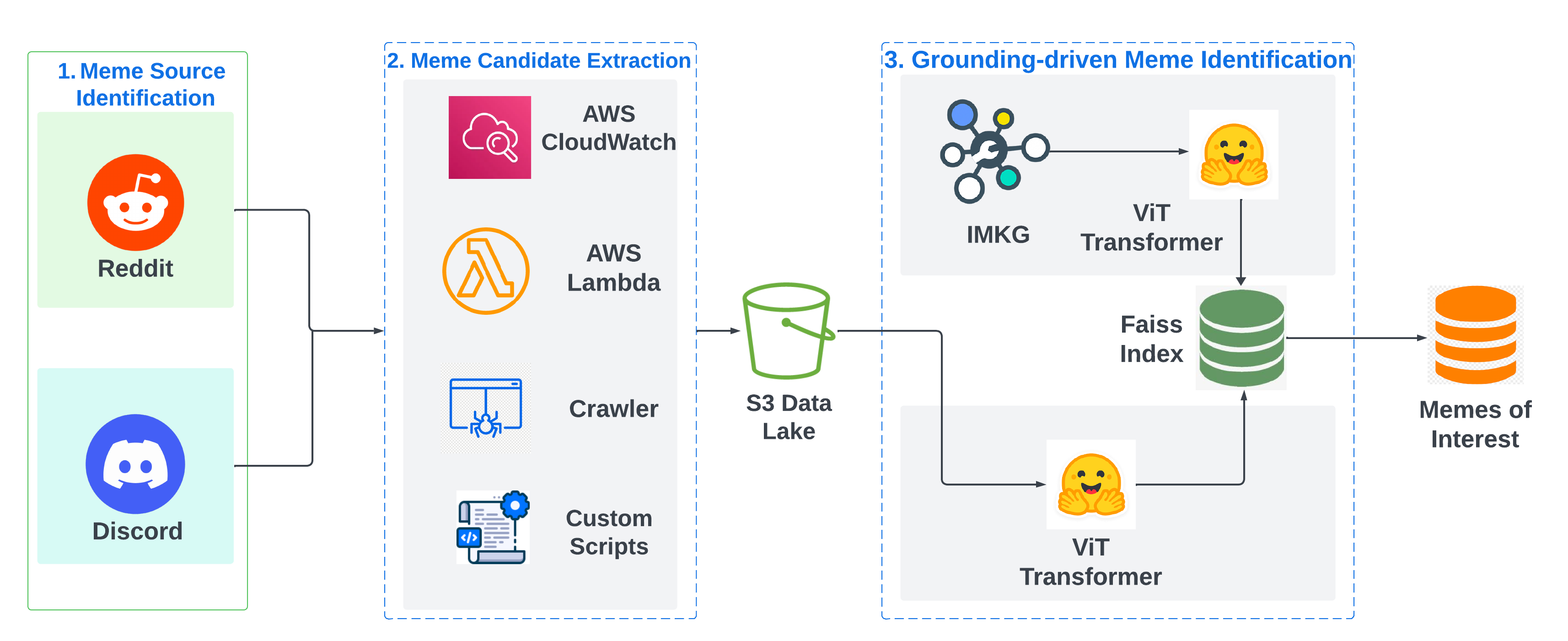}
    \caption{End-to-End Pipeline for IMs Contextualization empowered by IMKG}
    \label{fig:imkg_pipeline_process}
\end{figure*}

We develop a methodology motivated by the need to establish a precise and efficient method for meme identification on social media. 
Our methodological framework consists of the three sequential components presented in Figure \ref{fig:imkg_pipeline_process}. The \textit{Meme Source Identification} step singles out social media platforms and channels rich in meme content. Then, the \textit{Meme Candidate Extraction} process relies on the Extract, Transform, and Load (ETL) paradigm to extract raw data from these platforms and channels, refine it, and then load it into dedicated databases. The resulting data serves as a collection of meme candidates. Finally, the \textit{Grounding-driven Meme Identification} component performs filtering of these meme candidates, by matching them to a collection of known memes from IMKG based on visual similarity.


\subsection{Meme Source Identification}

Social media networks, such as Reddit, Twitter, and Discord, are essential sources for IM analysis given the remarkable pace and breadth at which creative communication spreads on these platforms~\cite{blommaert2015conviviality}. In the current version of our framework, we select two widely used social media platforms: Reddit and Discord. 
Reddit, with its vast user-generated content, serves as a primary hub for emerging memes~\cite{singer2014evolution}, while Discord, accommodating millions of global users, offers unrestricted access to meme-rich interactions within its diverse while goal-driven community servers~\cite{fairchild2020s,discord_groups}. Besides popularity, another criterion for prioritizing these two platforms is their data accessibility via APIs and established libraries.


\noindent \textbf{Reddit.}\footnote{https://www.reddit.com/} Reddit is a social media platform and online community that revolves around user-generated content and discussions. It was founded in 2005 and has grown to become one of the most popular websites on the internet. Much of the Internet's most popular content goes viral initially on Reddit, reflecting the website's reputation~\cite{barnes2021dank}. Reddit is organized into \textit{subreddits}, i.e., individual communities or forums dedicated to specific topics or interests. It provides a diverse array of user-generated content, including an extensive daily production of memes~\cite{Nieubuurt2021InternetML} with a wide range of themes, ideologies, and user demographics~\cite{doi:10.1177/1461444815608807}. 





We use PRAW,\footnote{https://github.com/praw-dev/praw} a Python Reddit API Wrapper, to access Reddit's API directly. This Python package enables us to retrieve posts from any subreddit. Reddit is an extensive and active community, which makes it challenging to capture its entirety, especially given the infrastructure constraints in place for site load balancing and throttling. To manage its availability, Reddit restricts the API's request frequency to no more than two requests per second. Additionally, the API has a return limit for a single request, ranging from 100 to 1500 objects depending on the object type and user level~\cite{10.1145/2567948.2579231}. To alleviate these challenges, we evaluated other tools such as Pushshift\footnote{https://github.com/pushshift/api} and PMAW\footnote{https://github.com/mattpodolak/pmaw}, the Pushshift Multithread API Wrapper; however, following Pushshift's discontinuation, we decided against using them. Ultimately, we opt for PRAW as the most suitable option with easy access, despite its inherent limitations. Each Reddit post in our dataset is characterized by the eleven features extracted through PRAW, namely \textit{title}, \textit{author}, \textit{selftext}, \textit{score}, \textit{ups}, \textit{downs}, \textit{created\_utc}, \textit{posturl}, \textit{\#comments}, \textit{imageurl}, and \textit{is\_nsfw}.

\noindent \textbf{Discord.}\footnote{\url{https://discord.com}} Discord is a popular community-driven communication platform, specifically catering to gamers and other interest groups. It was initially released in 2015 and has since gained widespread use for its text, voice, and video communication features. It hosts over 135 million global users, with 19 million users active daily. Discord enables community members within a server to communicate through forum posts, voice interactions, and sharing features reminiscent of social media platforms. Notably, many shared features are dominantly visual, including IMs, which are frequently reproduced and disseminated~\cite{fairchild2020s}. For meme collection from Discord, we use a custom package, designed specifically to extract chat messages from Discord servers. The resulting data contains seven features, namely \textit{content}, \textit{author}, \textit{pinned}, \textit{created\_utc}, \textit{mentions}, \textit{reactions}, and \textit{attachments}.


\subsection{Meme Candidate Extraction}

To generate meme candidates from the identified data sources, we develop an extract-transform-load (ETL) pipeline~\cite{vassiliadis2009survey}. 

\noindent \textbf{Extract.} To extract raw data from the target platforms identified in the previous step, an AWS CloudWatch alarm executes a custom script every 24 hours. This script, written using the AWS Python SDK, boto3,\footnote{https://boto3.amazonaws.com/v1/documentation/api/latest/index.html} fetches data by utilizing AWS Lambda's functionalities.
AWS Lambda is a serverless computing service that facilitates the execution of stateless functions in response to specific events~\cite{aws_lambda}. AWS Lambda's capability to support up to 3,000 simultaneous invocations ensures efficient data extraction. 

\noindent \textbf{Transform.} After extracting the data, we transform it to ensure that it aligns with our analytical objectives. Specifically, our transformation involves two main operations: retaining only images in jpg, jpeg, and png formats, and standardizing date stamps from various subreddits and channels to a consistent UTC format. We utilize both in-house scripts and AWS services to maintain data consistency. The result of the transformation step is thus a set of posts with images representing our meme candidates.

\noindent \textbf{Load.} After the transformation, we load the refined posts data into AWS S3, chosen for its scalability, availability, security, and performance~\cite{aws_s3}. Numerical, descriptive, and time attributes are stored in a designated S3 bucket, whereas images occupy a separate S3 repository, creating the foundation for our S3 Data Lake. This Data Lake provides a scalable and cost-effective storage environment~\cite{mathis2017data}, ideal for advanced analytics of extensive datasets and facilitating comprehensive meme analysis across platforms. In parallel, we employ AWS CloudWatch for real-time monitoring, using it to aggregate and display logs, metrics, and event data, ensuring efficient infrastructure and application maintenance~\cite{aws_cloudwatch}. 

\subsection{Grounding-driven Meme Identification}

\textbf{IMKG.} To detect memes among the collection of visual objects, we devise a method to ground them to IMKG.
IMKG is a recent knowledge graph of over 1.3 million memes corresponding to 1,765 unique templates, organized into three key classes: Media Frame, Template, and Meme instances. The Media Frame class objects encapsulate ``memeable'' situations, i.e., situations recognizable to both the meme creator and a wider audience. A Template embodies multimedia structures corresponding to a Media Frame, typically coupling an image with a caption text placeholder. These templates are common bases for the Meme instances, which are adaptations of the foundational template with myriad permutations. This model highlights their multimodality, brevity, relatability, and adaptability, and connects IMs to their original media frame and their underlying template. A snippet of IMKG can be found in \autoref{fig:meme_understanding} (green box).

The creation of IMKG encompasses four primary steps, starting with the crafting of a knowledge graph (KG) model that leverages the structured organization of Media Frame, Template, and Meme instances. Following this, data is amassed from key sources of internet memes, including encyclopedias, meme generation platforms, and extensive KGs. This data undergoes enhancement through textual and visual processing, supplemented by background knowledge, and is integrated in line with the initially designed model using a schema mapping language. The final IMKG is then made accessible in both RDF and labeled property graph (LPG) formats. Central to IMKG are Wikidata~\cite{vrandevcic2014wikidata}, serving as a foundational seed due to its item ``Internet meme (Q2927074)'', the KnowYourMeme (KYM) encyclopedia for meme lore, origin, and interpretation, and ImgFlip\footnote{\url{https://imgflip.com}}, offering insights into meme templates and user interactions. This integration aids in providing a holistic understanding of a meme's origins and variations, leveraging diverse characteristics such as template title, entities from caption, alternative text, and image, to enhance our grasp on the meme's architecture and context.

\textbf{Grounding memes to IMKG.} We leverage the organization of IMKG into Media Frames, Templates, and Meme Instances, and their semantic relationships, to systematically identify meme candidates from images sourced from Reddit and Discord. Namely, we select IM templates from IMKG, which are later used to discern social media memes from generic images. To interact with IMKG and extract relevant meme templates along with their associated instances, we utilize Kypher—a query language for hyperrelational graphs.\footnote{\url{https://kgtk.readthedocs.io/en/latest/transform/query/}}
Then, we leverage a visual encoder, the Vision Transformer (ViT) model~\cite{dosovitskiy2021image}, to align memes from social media to the templates extracted from IMKG.
Unlike traditional convolutional neural networks that exploit local spatial hierarchies~\cite{o2015introduction}, ViT leverages the Transformer architecture, predominantly used for natural language processing~\cite{vaswani2017attention}. In the ViT model, an image is divided into fixed-size patches, which are then linearly embedded to form a sequence of flat vectors. To retain the spatial context, positional embeddings are incorporated. These embedded sequences serve as input to the Transformer, ensuring global image understanding. The importance of ViT stems from its ability to scale effectively with increased data and model size, often outperforming state-of-the-art CNNs when ample data is available~\cite{chen2021vision}. 
As multiple meme instances of the same template have different representations, we encode three images from each IMKG template and we utilize FAISS for vector similarity search~\cite{johnson2019billion}. In parallel, we encode images sourced from platforms like Reddit and Discord, housed in our S3 Data Lake, using the same ViT model. The encoded images are then used as queries against the FAISS index to determine whether they are memes or not. Namely, the images from Reddit and Discord based on their similarity scores against the FAISS index are evaluated with an experimentally determined \textit{threshold} $t$, and those falling below $t$ are classified as non-memes. Conversely, images whose similarity surpasses $t$ are designated as \textit{memes of interest}.

\section{Experimental Setup}
\label{sec:setup}

\begin{figure*}[ht!]
    \centering
    \includegraphics[width=0.7\textwidth]{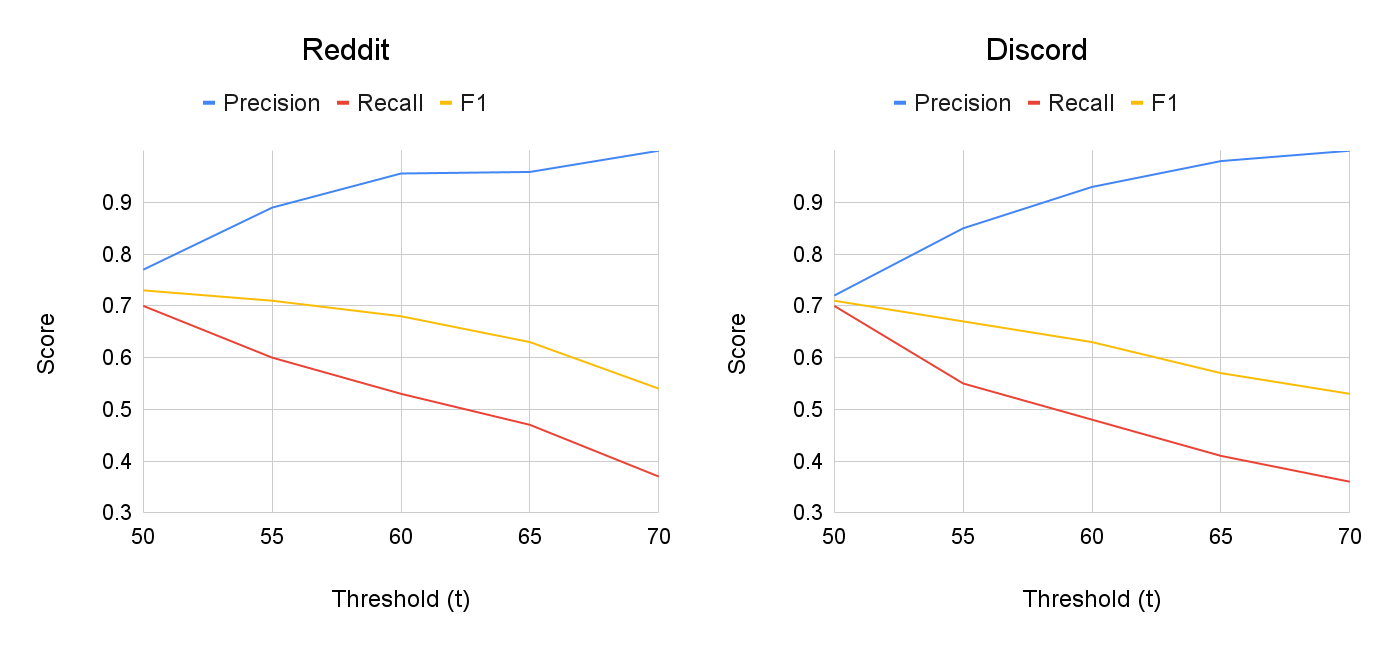}
    \caption{Performance metrics across varying threshold values.}
    \label{fig:rq_m1}
\end{figure*}

\begin{figure}[h]
    \centering
    \includegraphics[width=0.7\linewidth]{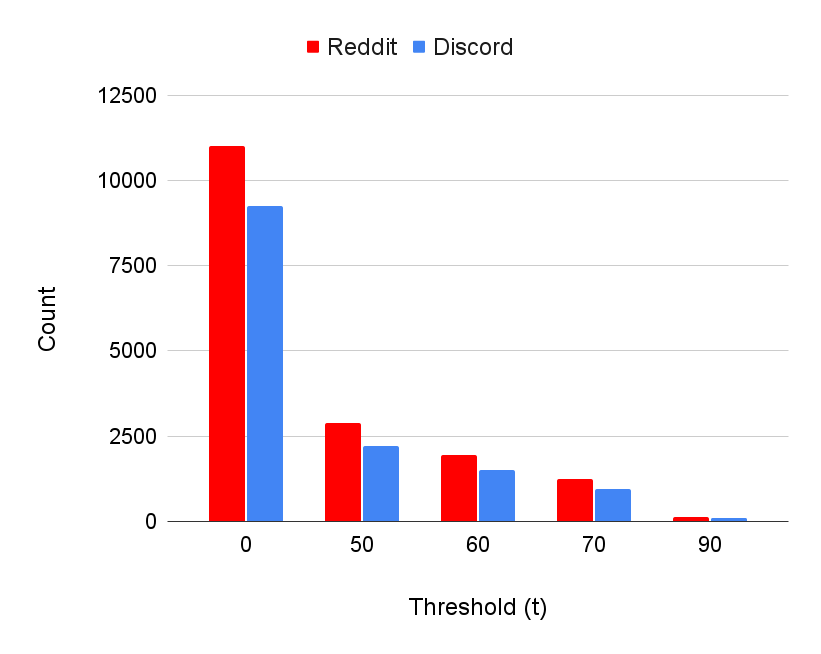}
    \caption{Meme distribution across threshold values.}
    \label{fig:filter_comparison}
\end{figure}

\textbf{Data selection} We determine our meme collection strategy by analyzing which channels and subreddits had the most memes based on a preliminary analysis conducted during May and June of 2023. We initially selected 15 Reddit subreddits and 15 Discord channels based on a manual investigation aimed at selecting online communities diffusing a high volume of IMs, i.e., the set of subreddits and Discord channels are chosen for the active participation of their communities in the meme culture. The candidate Reddit subreddits and Discord channels are listed in the \textit{Appendix} (see Table \ref{tab:preliminary_data}). Next, we leverage the pipeline detailed in Section \ref{sec:method} to analyze the meme content in 1,000 posts from each of these communities, as a preliminary analysis. This analysis identified five subreddits - \texttt{r/memes}, \texttt{r/meme}, \texttt{r/HistoryMemes}, \texttt{r/PoliticalMemes}, and \texttt{r/ProgrammerHumor}, as well as five Discord channels - \texttt{auto\_memes\_2}, \texttt{TheDungeon}, \texttt{meme\_shitposting}, \texttt{MemeStash}, and \texttt{meme\_stealing}, with the highest volume of shared meme content. According to our preliminary analysis, on average the selected five subreddits contained 130 meme posts, in contrast to the 38 meme posts in the non-selected subreddits. Similarly, the chosen Discord channels averaged 120 meme posts compared to 17 in the channels not considered for final evaluation. Based on these insights, we perform the final analysis of these ten communities over one month, starting on July 1st and ending on July 31st of 2023. By doing so, we ensure the relevance and timeliness of our data and a fair comparison between the two media platforms.

\noindent \textbf{Modeling details} We use the google/vit-base-patch16-224-in21k model from HuggingFace~\cite{wu2020visual} to encode images from IMKG template memes as well as those from Reddit and Discord. This model is a Transformer encoder \cite{vaswani2017attention} pre-trained on ImageNet-21k~\cite{deng2009imagenet} at a 224x224 pixel resolution. The images from our meme collection are segmented into 16x16 patches and linearly embedded. A [CLS] token is appended alongside absolute position embeddings to the linearly embedded image patches, and then these enriched representations are processed by the Transformer encoder. Upon encoding the image with the Transformer encoder, the embedding vector for a meme is extracted from the final Transformer layer's [CLS] token representations. The FAISS index is configured to utilize a Flat configuration, guaranteeing that an exhaustive, brute-force search is executed across all stored vectors during query operations, with the inner product serving as the similarity metric. The vectors are subjected to a normalization process before their addition to the index to ensure they are of unit length. This normalization ensures that the inner product between vectors equates to the cosine similarity, enhancing the reliability of similarity computations. 

\begin{table*}
    \centering
    \caption{Distribution of \textbf{P}osts, \textbf{I}mage \textbf{P}osts, \textbf{M}eme \textbf{P}osts, and \textbf{M}eme to \textbf{I}mage \textbf{R}atio in Reddit subreddits and Discord channels.}
    \label{tab:image_stats_table}
    \small
    \begin{tabular}{|c|c|c|c|c|c|c|c|c|c|}
        \hline
        \multicolumn{5}{|c|}{\bf Reddit} & \multicolumn{5}{c|}{\bf Discord} \\ \hline
        \bf Subreddit & \bf \#P & \bf \#IP & \bf \#MP & \bf \#MIR & \bf Channel & \bf \#P & \bf \#IP & \bf \#MP & \bf \#MIR \\ \hline
        r/meme   & 2794   & 2243 & 392  & 0.17 & auto\_memes\_2   & 1679   & 1341  & 252 & 0.18 \\ \hline
        r/PoliticalMemes   & 716   & 653 & 53  & 0.08 & meme\_shitposting  & 926  & 611 & 22 & 0.03 \\ \hline
        r/HistoryMemes   & 1635   & 1414 & 237 & 0.16 & TheDungeon  & 12044  & 6624 & 1184 & 0.17 \\ \hline
        r/ProgrammerHumor   & 901   & 626 & 151 & 0.24 & meme\_stealing  & 205  & 177 & 3 & 0.01  \\ \hline
        r/memes   & 7129   & 6561 & 1120  & 0.17 & MemeStash  & 1131  & 1072 & 33 & 0.03 \\ \hline
        \bf Total   & \bf  13175   & \bf 11497 & \bf 1953  & \bf 0.16  & \bf Total  & \bf 15985  & \bf 9825 & \bf 1494   & \bf 0.15 \\ \hline
    \end{tabular}
\end{table*}

\begin{figure*}[ht!]
    \centering
    \includegraphics[width=0.7\textwidth]{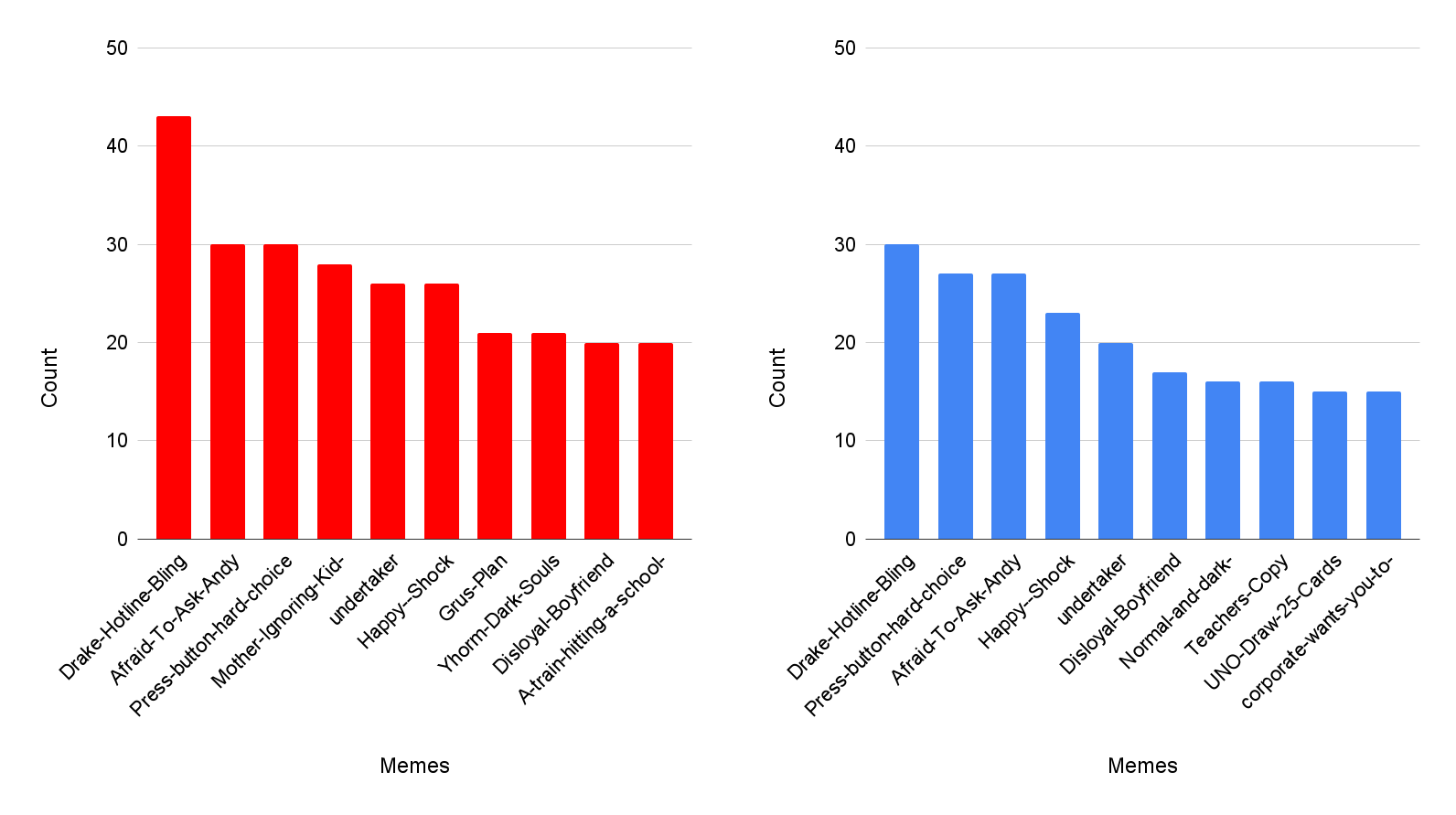}
    \caption{Comparative distribution of the most popular IMKG memes on Reddit and Discord.}
    \label{fig:top_memes}
\end{figure*}

\section{Results}
\label{sec:results}

\subsection{RQ1: Mapping social media memes to IMKG}

To address RQ1, we investigate the possibility of mapping memes from Reddit and Discord to IMKG and the impact of the similarity threshold $t$ on the precision and recall of meme recognition. We experiment with thresholds in the range between 50\% to 70\%. For our analysis, a meme is defined based on its adherence to recognized meme templates and its presence within the IMKG database, reflecting both visual and conceptual alignment with established meme formats. To evaluate the efficiency and precision of these filters, we manually annotate a dataset comprised of 200 randomly chosen posts (92 memes and 108 non-memes). This allows us to determine the optimal filter to maximize both precision (the proportion of correctly identified memes out of all posts identified as memes) and recall (the proportion of actual memes correctly identified out of all memes). Figure \ref{fig:rq_m1} shows that using a 70\% threshold, our model reaches perfect precision, which means that every image identified as a meme is indeed a meme. However, the perfect precision is accompanied by a relatively low recall (under 0.4), revealing that $t=70\%$ would likely overlook a significant portion of the meme posts. In contrast, the 50\% threshold, although capturing a wider range of meme content and having the optimal precision-recall tradeoff, inadvertently introduces a higher rate of false positives, misidentifying many non-memes as memes. We consider precision to be more important than recall for our study to have a high-quality analysis. Thus, we pick the threshold with the highest recall possible at a precision over 0.9, which is 60\%. The 60\% threshold yields precision scores of 0.95 and 0.93, respectively for Reddit and Discord; meanwhile, the precision values at the 50\% threshold are 0.77 and 0.72. In Figure \ref{fig:filter_comparison}, we show the number of images classified as memes using various threshold values - 50\%, 60\%, 70\%, 90\%. By applying the 60\% threshold to our Reddit and Discord datasets, we adeptly filter out non-memes, resulting in a curated database abundant in target memes, which we refer to as \textit{memes of interest}.  With this selected threshold, we process our dataset from Reddit and Discord. In Figure \ref{fig:memes_f1} in the Appendix, we present examples of our classification results.


\subsection{RQ2:  Prevalence of IMKG memes on Reddit and Discord}


The distribution of the selected subreddit and Discord channel posts, as well as those with an image (i.e., \textit{image posts}), is presented in Table \ref{tab:image_stats_table}. Specifically, the total number of posts from Reddit is 13,175, of which 11,497 (87\%) are identified as \textit{image posts}. Similarly, for Discord, the total number of posts is 15,985, and 9,825 (61\%) of which are \textit{image posts}. Using a 60\% threshold, 1,953 images from Reddit (17.75\%) and 1,494 from Discord (representing 16.18\%) align with the IMKG database. 
In terms of specific communities, we see that over half of the memes in our Reddit collection come from the subreddit \texttt{r/memes}, and over half of the Discord memes come from the channel \texttt{TheDungeon}. These numbers, given the vast meme diversity on these platforms, show IMKG's promise as an instrumental tool in internet meme research.
The highest relative yield of memes is observed for the \texttt{r/ProgrammerHumor} subreddit, for which 24\% of the images are identified as memes.
Originating from distinct subreddits on Reddit and varied channels on Discord, this refined collection from a one-month original data is already rich in meme content, with an estimated precision of over 90\%.

\begin{figure}[h!]
    \centering
    \includegraphics[width=0.5\textwidth]{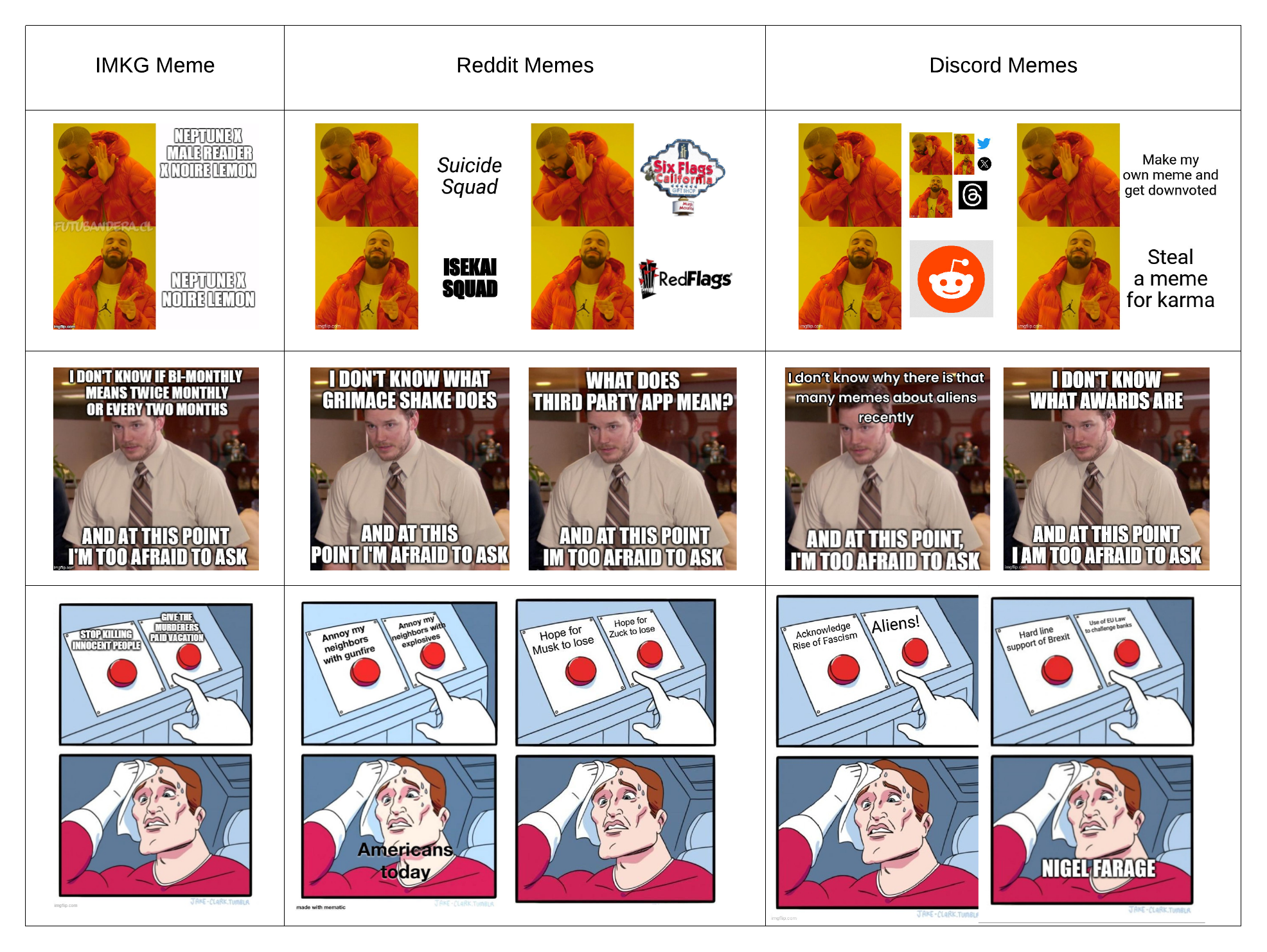}
    \caption{Mapping between IMKG, Reddit, and Discord for three popular memes: \textit{Drake-Hotline-Bling}, \textit{Afraid-To-Ask-Andy}, and \textit{Press-button-hard-choice}.}
    \label{fig:results2_memes}
\end{figure}

\subsection{RQ3: Popularity of IMKG memes on Reddit and Discord}

\begin{figure*}[ht!]
    \centering
    \includegraphics[width=0.63\textwidth]{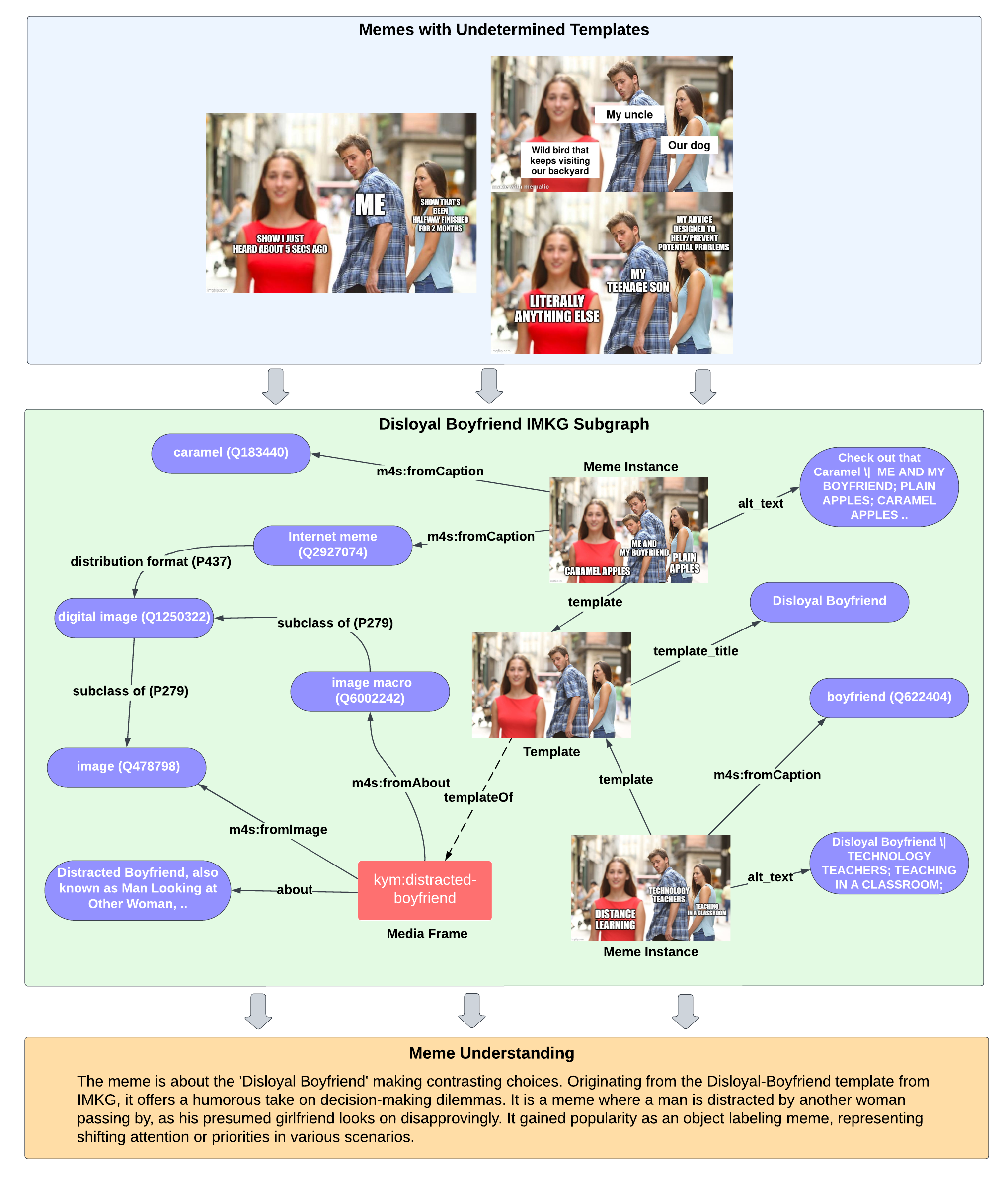}
    \caption{IM contextualization using the IMKG knowledge, illustrated on the Disloyal Boyfriend meme.}
    \label{fig:meme_understanding}
\end{figure*}

Determining the most trending IMKG meme templates within the target online communities is important as it provides a glimpse into the digital meme culture, revealing time-dependent user preferences and shared humor across platforms. The popularity of these memes is calculated based on the number of occurrences of posts across platforms, allowing for a quantitative measure of their prevalence and the extent to which they resonate within the community.
Accordingly, Figure \ref{fig:top_memes} provides an overview of the most popular meme templates on both Reddit and Discord. From this visualization, we observe that the \textit{Drake-Hotline-Bling} meme was top-rated on both Reddit and Discord. Also, 4 of the top 5 memes, namely \textit{Drake-Hotline-Bling}, \textit{Afraid-To-Ask-Andy}, \textit{Press-button-hard-choice}, and \textit{undertaker}, were common on both platforms. These popular memes show what has been trending in the digital meme world in July 2023, which can be useful for marketers, researchers, and regular users. For three of these memes, Figure \ref{fig:results2_memes} provides examples of two of their instances from both Reddit and Discord, as well as an example corresponding meme found in IMKG. This figure illustrates the relatability of these memes (\textit{Drake-Hotline-Bling}, \textit{Afraid-To-Ask-Andy}, and \textit{Press-button-hard-choice}), as they have been adapted to refer to a wide variety of situations expressing the same core idea. For example, the \textit{Afraid-To-Ask-Andy} meme is used to refer to the linguistic ambiguity of the term ``bi-monthly'' as well as the prevalence of alien memes, as they are both memes that express a recognizable confusion about a certain phenomenon.  Our choice to match memes based on visual features is underpinned by the objective of bridging the gap between different instances of the same meme template, highlighting the universal appeal and adaptability of these visual jokes. This approach allows us to connect memes that, despite having varied textual content, share similar visual structures, thereby maintaining the integrity of the meme's core message across diverse platforms. We note how IMKG serves as an anchor for these memes to be grounded from such different platforms, which is enabled by this strategic matching process, thus connecting memes regardless of their underlying ideas.

\subsection{RQ4: Enhancing meme understanding with IMKG}

In this case study, we illustrate the potential of IMKG as a unique approach to enhancing the understanding of memes, especially those without defined templates. Specifically, we examine the ``Disloyal Boyfriend'' memes sourced from platforms such as Reddit and Discord utilizing IMKG, emphasizing its utility in structuring and understanding these memes. Through Figure \ref{fig:meme_understanding}, it becomes evident that the template, represented as imgflipmeme:112006116/Disloyal-Boyfriend, is connected to both the original Media Frame as well as numerous meme instances. Through the media frame, we access encyclopedic information from KnowYourMeme present in IMKG, such as a general ``about'' description, ``origin'' as an exhaustive chronicle of the meme's inception and evolution, and tags like ``disloyal man with his girlfriend looking at another girl''. The integration of this data into IMKG, from sources like KnowYourMeme and ImgFlip, along with Wikidata entities extracted from the image (\texttt{m4s:fromImage}), from the instance captions (\texttt{m4s:fromCaption}), and from the description (\texttt{m4s:fromAbout}), demonstrates the structured approach of IMKG in connecting memes to a broad spectrum of world knowledge. The meme instances are associated with several pieces of information, including their alternative text for easy access to the caption. This direct linkage between meme templates, their originating contexts, and subsequent variations underscores IMKG's comprehensive framework for meme analysis.

While IMKG provides information to contextualize social media memes in terms of existing memes and world knowledge, we also note three drawbacks of its current version. First, the dotted line connecting the template and the media frame in the figure underscores a standing challenge of entity resolution in IMKG, where certain templates are not formally linked to their media frame. Second, while we note that IMKG provides useful information highlighted through our manual summary in the yellow box for supporting social media meme interpretation, this detailed interpretation emerges from our focused examination of IMKG's structure rather than being an inherent feature of the graph. Specifically, this involves selectively examining relevant subgraphs within IMKG to facilitate a more manageable and insightful analysis. This approach underscores the necessity of navigating IMKG's complexity through targeted knowledge selection, which is critical for distilling meaningful insights from the vast array of data. Such a targeted exploration is essential for effectively utilizing IMKG to comprehend the multifaceted nature of memes, their cultural significance, and their evolution over time. Third, the dynamic nature of meme culture, with new memes emerging daily, presents an additional challenge, as these will not be immediately present in IMKG, potentially limiting the methodology's ability to capture the most recent trends. These caveats are key challenges that will be addressed in future work to make IMKG more effective for IM contextualization.




\section{Conclusions}
\label{sec:conclusion}

In this work, we introduced and assessed a novel methodological framework for identifying and characterizing IMs from social media posts by mapping them to IMKG. Our approach harnessed data from popular accessible social media platforms, Reddit and Discord, enabling the identification, categorization, and subsequent analysis of IMs within these ecosystems. Leveraging Vision Transformers for encoding and visual matching, our pipeline grounded memes from diverse social platforms to IMKG. By following our procedure, we identified nearly 3,500 memes from Reddit and Discord with a precision close to 90\%. Our analysis of these IMs yielded a set of common meme communities and revealed that the most popular IMs were shared across the two platforms. 
Finally, we performed a case study to understand how IMKG grounding enables IM contextualization through its encyclopedic information.

Our findings underscore the potential and importance of utilizing a semantic repository for decoding the evolving IM landscape, and open research directions for identity-driven analytics of memes, expanding to fringe platforms, and new methods for hate speech detection on the Web. Meanwhile, a key limitation of our approach is that it does not incorporate textual or other contextual information present in IMKG and also does not address potential biases in meme selection or the impact of evolving meme formats on our analysis. We anticipate that our work will inspire further interdisciplinary research at the intersection of social media and KGs, e.g., 
to investigate how many IMs go undetected because they are missing in IMKG, which might inform further work on completing IMKG.

\bibliographystyle{ACM-Reference-Format}
\bibliography{sample-base}

\appendix
\section{Appendix}

In Table \ref{tab:preliminary_data}, we show the subreddits and channels we selected for our preliminary analysis on IMs use on Reddit and Discord, respectively.

\begin{figure}[h!]
    \centering
    \includegraphics[width=0.5\textwidth]{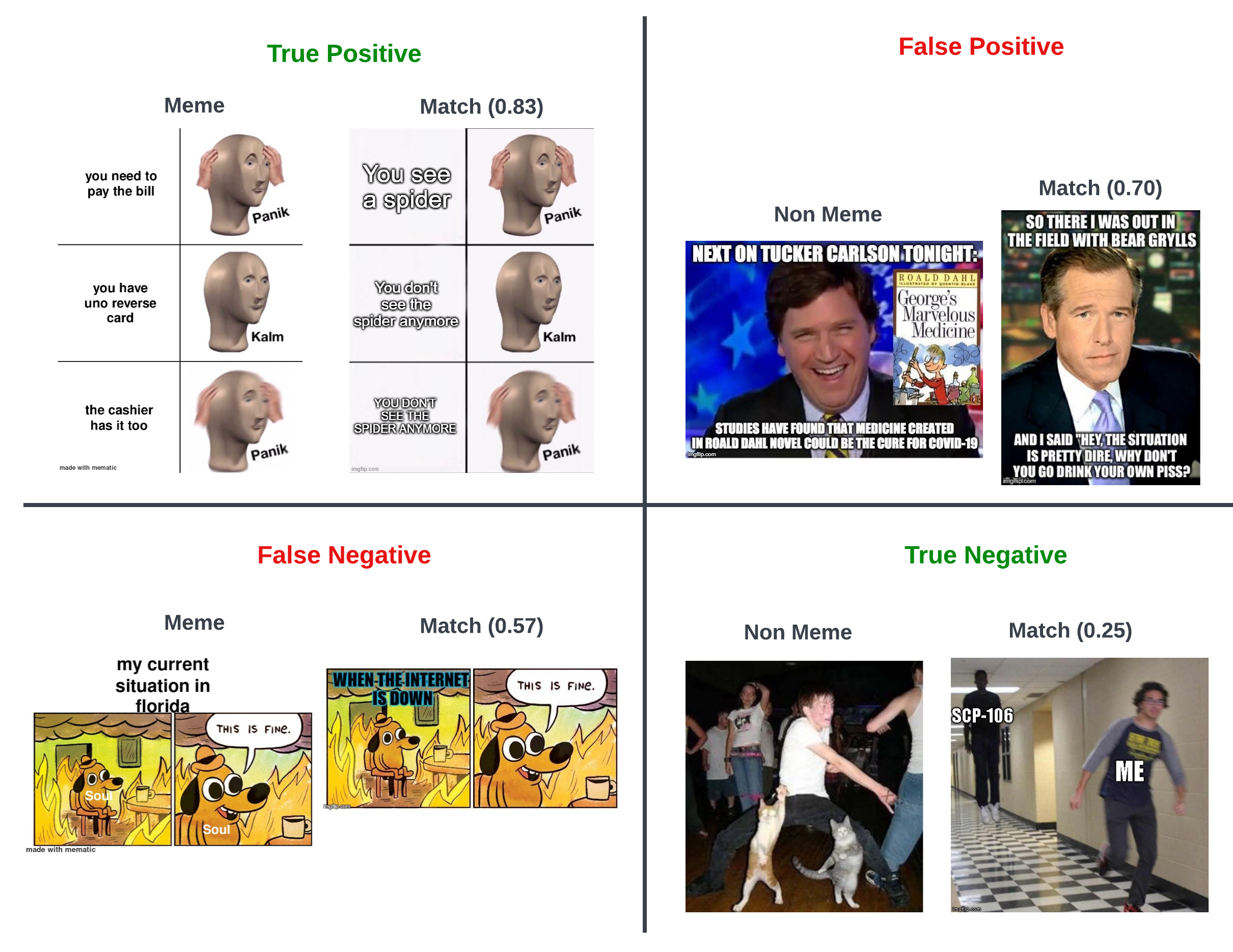}
    \caption{Classification results using a 60\% threshold.}
    \label{fig:memes_f1}
\end{figure}

In Figure \ref{fig:memes_f1}, we present examples of our classification results for true positives, false positives, true negatives, and false negatives. Each quadrant of the figure showcases a pair of images: one sourced from a social network (Reddit or Discord), and its corresponding match (or mismatch) from the IMKG. Accompanying these image pairs, we have incorporated a match score, which quantitatively reflects the degree of similarity between the images. Scores exceeding the threshold of 60\% are deemed positive, while those below it are treated as negative. On the one hand, this figure shows that true positives and true negatives can be distinguished easily (quadrants 1 and 4). Meanwhile, the presence of false positives and false negatives (quadrants 2 and 3) underscores the inherent complexity in aligning memes from various platforms with the IMKG, 
attributed to the mutable and multimodal nature of memes, which often intertwine nuanced visual and textual elements, sarcasm, and context-dependent humor. This complexity highlights the challenges in meme identification, which are not entirely solved by tuning the threshold parameter.

\begin{table}[h!]
    \centering
    \caption{Selected Reddit subreddits and Discord channels for preliminary meme content analysis.}
    \label{tab:preliminary_data}
    \begin{tabular}{|c|c|}
        \hline
        \textbf{Subreddit} & \textbf{Discord Channel} \\
        \hline
        r/TheLeftCantMeme & auto\_memes\_2 \\
        \hline
        r/meme & cozy\_lounge \\
        \hline
        r/wholesomememes & DankersContinental \\
        \hline
        r/HistoryMemes & MemeCentral \\
        \hline
        r/memes & memee \\
        \hline
        r/dank\_meme & memeplex \\
        \hline
        r/PreqelMemes & memes\_boss \\
        \hline
        r/funny & meme\_shitposting \\
        \hline
        r/CoronaVirusMemes & MemeStash \\
        \hline
        r/ProgrammerHumor & meme\_stealing \\
        \hline
        r/dankmeme & memeverse \\
        \hline
        r/PoliticalMemes & memex \\
        \hline
        r/MemeEconomy & Socio \\
        \hline
        r/TheRightCantMeme & TheDungeon \\
        \hline
        r/me\_irl & WeebCommunity \\
        \hline
    \end{tabular}
\end{table}


\end{document}